\documentstyle[12pt,epsfig]{article}

\begin{document}
\newcounter{num}
\def\lsim{\mathrel{\lower4pt\hbox{$\sim$}}\hskip-12pt\raise1.6pt\hbox{$<$}\;}
\def\BAR{\bar}
\def\fm{{\cal M}}
\def\fl{{\cal L}}
\def\fp{{\cal P}}
\def\n{{n}}
\def\delt{{n}}
\def\gsim{\mathrel{\lower4pt\hbox{$\sim$}}
\hskip-10pt\raise1.6pt\hbox{$>$}\;}

\vspace*{-.5in}
\rightline{AMES-HET 9909}   
\rightline{ROME1-1277/99}
\rightline{BNL-HET-99/79}
\begin{center}

{\large\bf Dijet Production at Hadron Colliders in  Theories with Large 
Extra Dimensions}\\

\vspace{.3in}

David Atwood$^{1}$\\ 
\noindent Department of Physics and Astronomy, Iowa State University, Ames,
IA\ \ \hspace*{6pt}50011\\
\medskip

Shaouly Bar-Shalom$^{2}$\\ 
\noindent  
INFN, Sezione di Roma 1 and Physics Department, 
University ``La Sapienza'', Rome, Italy
\\
\medskip

\medskip
and\\
\medskip

Amarjit Soni$^{3}$\\
\noindent Theory Group, Brookhaven National Laboratory, Upton, NY\ \ 
11973\\

\footnotetext[1]{email: atwood@iastate.edu}
\footnotetext[2]{email:  Shaouly.BarShalom@Roma1.infn.it}
\footnotetext[3]{email: soni@bnl.gov}
\end{center}
\vspace{.2in}

\begin{quote}
{\bf Abstract:}

We consider the production of high invariant mass jet pairs at hadron
colliders as a test for TeV scale gravitational effects. We find that this
signal can probe effective Planck masses of about $10~TeV$ at the LHC with
center of mass energy of $14~TeV$ and $1.5~TeV$ at the Tevatron with
center of mass energy of $2~TeV$. These results are compared to analogous
scattering processes at leptonic colliders.

\end{quote} 
\vspace{.3in} 
\newpage

Gravitation is by far the weakest force of nature. Indeed, the usual
explanation of this is that quantum gravitational effect only become
important at the Planck mass scale $M_P=G_N^{-1/2}=1.22\times
10^{19}~GeV$. The fact that this scale is so much higher than the Standard
Model (SM) electroweak scale of $O(100~GeV)$ leads to the hierarchy
problem that fine tuning of the parameters of the SM at the Planck scale
are required to keep the electroweak scale small.  In general this may be
solved by supposing that new physical processes, such as SUSY or new
strong interactions, become manifest at the TeV scale.  Such new phenomena
can thus lead to a low energy effective theory that does not depend on the 
exact parameters of physics at the Planck scale.

Many string theories such as M-theory~\cite{mtheory} can only be
consistent if there are more than 3+1 dimensions, presumably the extra
dimensions would form a compact manifold. This leads to the recent
suggestion of~\cite{add1,add2} that gravity, which is weak at macroscopic
scales, but may become strong at $TeV$ scales. In particular, if there are
$n$ compact dimensions of length $R$, at distances $d<R$ the Newtonian
inverse square law will fail~\cite{add1} and the gravitational force will
grow at a rate of $1/d^{n+2}$. If $R$ is sufficiently large, even the weak
strength of gravitational force at the macroscopic scale can lead to a
strong force at distances of $1~TeV^{-1}$. For this to happen, the size of
the extra dimension should be: 

\begin{eqnarray}
8\pi R^n M_S^{2+n}\sim M_P^2 ~.
\label{Rsize}
\end{eqnarray}

\noindent where $M_S$ is the effective Planck scale of the theory which is
the mass scale at which quantum gravitational effects become important.
There is no hierarchy problem if the scale $M_S$ is $O(1~TeV)$ i.e. not 
far beyond the electroweak scale. 
For instance, if $n=1$ and $M_S$=$1~TeV$, then R is of the order of
$10^{8}~km$, large on the scale of the solar system and clearly ruled out
by astronomical observations. However, if $\delt\geq 2$ and $M_S\geq
1~TeV$ then $R< 1~mm$; there are no experimental constraints on the
behavior of gravitation at such distance scales~\cite{cavin}. This
compactification is thus not ruled out based on gravitational experiments
(there are however alternative schemes which can consistently allow one
extra dimension such as in~\cite{randall}).

Of course all other forces and particles appear to exist in the usual
$3+1$ dimensions. In the proposed scenario of~\cite{add1} this results
from the existence of a 3+1 dimensional brane to which all known fermions
and gauge fields are confined in the total of 3+n+1 dimensional space.
Only gravitation can propagate through the bulk and therefore may directly
be sensitive to the effects of the new dimensions.

Thus, gravitational effects can become important at $TeV$ scale colliders.
In particular, the onset of strong gravitational effects may be understood
by perturbative couplings of matter to gravitons leading to observable
effects due to the production of real graviton states or the exchange of
virtual gravitons at energies approaching $M_S$.

To calculate such perturbative gravitational effects, it is useful to
adopt the 4-dimensional point of view. We therefore interpret the graviton
states which move parallel to the 4 dimensions of space time as the usual
massless graviton giving rise to Newtonian gravity, while the graviton
states with momentum components perpendicular to the brane are effectively
a continuum of massive objects.  The density of graviton states is given
by~\cite{add1,add2,wells,taohan}:

\begin{eqnarray}
\rho(m^2)={dN\over d m^2}=
{m^{\delt-2}\over G_N M_S^{\delt+2}}
\end{eqnarray}

\noindent
where $m$ is the mass of the graviton.

Furthermore, gravitons with polarizations that lie entirely within the
physical dimensions are effective spin 2 objects.  Gravitons with
polarizations partially or completely perpendicular to the physical brane
are vector and scalar objects. In this letter, we will primarily be
concerned with the effects of the exchange of virtual spin 2 gravitons. To
perform perturbative calculations in this theory, one can formulate
Feynman rules for the coupling of graviton states to ordinary particles
where $\kappa=\sqrt{16 \pi G_N}$ is the effective expansion
parameter~\cite{wells,taohan}, in particular, we adopt the conventions
of~\cite{taohan}.

In the case of the exchange of virtual graviton states, one must add
coherently the effect of each graviton. For instance, in the case of
an $s$-channel exchange, the propagator is proportional to
$i/(s-m_{G_\lambda}^2)$ where $m_{G_\lambda}$ is the mass of the graviton
state $G_\lambda$. Thus, when the effects of all the gravitons are taken
together, the amplitude is proportional to

\begin{eqnarray}
\sum_\lambda {i\over s-m_{G_\lambda}^2}=D(s).
\end{eqnarray}

\noindent If $n\geq 2$ this sum is formally divergent as $m_{G_\lambda}$
becomes large. 
We assume that the distribution has a
cutoff at $m_{G_\lambda}\sim M_S$, where the underlying theory becomes
manifest. Taking this point of view, the value of $D(s)$ is calculated
in~\cite{wells,taohan}: 

\begin{eqnarray}
\kappa^2 D(s)=
-i{16\pi\over M_S^4}F + O({s\over M_S^2}) ~.
\end{eqnarray}

\noindent
The quantity $F$ contains all the dependence on $n$ and is given by:

\begin{eqnarray}
F=\left \{
\begin{array}{cl}
\log(s/M_S^2)    & ~~~{\rm for}~n=2\\
2/(n-2)          & ~~~{\rm for}~n>2
\end{array}
\right .
\label{eqf}.
\end{eqnarray}

In a $2\to 2$ process, a similar expression will apply for $t$ and $u$
channel exchanges.  If $n>2$, $D(s)$ is independent of $s$ in this
approximation and likewise the sum of the propagators in the $t$ and $u$
channels will be identical. As pointed out in~\cite{hooman}, this will not
necessarily be a good approximation in the case of $n=2$ because of the
logarithmic dependence of $D$ on $s$.

The theory formulated in this way does not treat the cutoff in detail but 
makes the ad hoc assumption that the cutoff is $M_S$. However, bounds 
which are obtained in this way may be applied to a more specific theory 
by computing an effective $M_S$ which would follow from the parameters of 
a given theory.
We can thus investigate the phenomenology which may occur
at various colliders~\cite{hooman}-\cite{catch} as well as precision
experiments~\cite{precision}. The assumption that the cutoff is $O(M_S)$
may be realized in a natural way from recoil effects of the brane as
discussed in~\cite{bando}, which gives rise to an exponential cutoff in the
coupling to gravitons with a mass greater than the stiffness of the brane.
In general the theory may be cut off by whatever new physical processes 
become manifest at $M_S$.

In attempting to place limits on such theories at a hadronic collider, the
most natural process to consider is one which produces real gravitons. 
For instance, if such gravitons were produced in association with a jet,
the monojet + large missing $P_T$ signal should be unmistakable.  Indeed
this process was considered in~\cite{monojet} and it was found that a
bound of $M_S=1.3$, $0.9$, $0.8~TeV$ may eventually be achieved at the
Tevatron for $n=2$, $4$ and $6$, respectively. At the LHC these bounds may
may be extended to $M_S=4.5$, $3.4$ and $3.3~TeV$ for $n=2$, $4$ and $6$.
The analogous process at the NLC, $e^+e^-\to \gamma G$ ($G=$graviton), was
also considered in~\cite{monojet,Cheung:1999zw} giving a reach at
$\sqrt{s}=1~TeV$ of $M_S=7.7$, $4.5$ and $3.1~TeV$ for $n=2$, $4$ and $6$.
Slightly better bounds may be obtained in the case of an $e\gamma$
collider based on a 1~TeV $e^+e^-$ collider where the reaction would be
$e\gamma\to eG$~\cite{e_gamma}.

For such processes which produce real gravitons, the cross section is
proportional to $(E/M_S)^{n+2}$ so that at larger $n$ less stringent
bounds can be placed on $M_S$. At $n=2$, there are also astrophysical
constraints, both from the rate of supernova
cooling~\cite{ahdd_astro,astro_star} which requires that $M_S>30~TeV$, and
the absence of a diffuse cosmic gamma ray background from the decay of
relic gravitons~\cite{ahdd_astro,astro_univ}, which requires $M_S>130~TeV$.
This latter bound, however, depends strongly on the assumption that all the 
decay modes of the graviton are governed by the perturbative decay modes.

While real graviton production bounds $M_S$ most stringently in the case 
where $n=2$, virtual processes tend to give better bounds in the case 
where $n>2$. 
In particular, we  
consider the observation of gravitational effects in 2 jet
events at hadron colliders, either $pp\to 2~jets+X$ or $p\bar p\to
2~jets+X$. As can be seen from eq.~(\ref{eqf}) the bounds obtained in
virtual graviton exchange events will be relatively independent of $n$.

It should also be kept in mind that the TeV scale gravitational theories
imply the existence of new physics at the scale of $M_S$ which may also
lead to two jet processes, such as discussed in~\cite{otherop2j}. Thus, in
experimentally probing the two jet signal, or indeed any manifestation of
virtual graviton exchange, one can only place limits on the gravitational
effects common to all such models. If a signal is seen, of course, further
investigation and observations in other channels is required to determine
if the effects are purely gravitational or if other physics is the cause.
Here we shall confine ourselves to a discussion of the limit that can be
placed on the effective $M_S$ from two jet events generated by graviton
exchanges.

At the parton level, two jet events are generated via processes of the
form $\rho_1\rho_2\to\rho_3\rho_4$. where $\rho_\ell$ are partons (of
momentum $p_\ell$). In particular, the possible parton level processes are
as follows:

\begin{eqnarray}
\begin{array}{lll}
(a)~q\bar q\to q^\prime \bar q^\prime &
(b)~qq^\prime \to q  q^\prime~/~q\bar q^\prime \to q \bar q^\prime&
(c)~qq\to qq \\
(d)~q \bar q \to q \bar q &
(e)~gg\to q \bar q / q \bar q \to gg  &
(f)~gq\to gq~/~g\bar q\to g\bar q\\
(g)~gg\to gg, & & 
\end{array}
\end{eqnarray}

\noindent where $q$ represents some flavor of quark and $q^\prime\neq q$ 
is a distinct flavor.

Of course each of these scattering processes has a
SM contribution which the gravitational amplitudes will
interfere with (where allowed by color conservation). 
We shall see however that since the amplitude grows with
$s^2$, scattering through gravitons tends to be harder and is easily
separated from SM processes which drop with $s$. 

The tree-level hard cross-sections $\sigma_i$ for a given 
subprocesses $i$, including the gravitational effects and their 
interference with the SM, can be written in the form:

\begin{eqnarray}
{d\sigma_i\over d z}
=
k_s\left[
{\pi \alpha_s^2\over 2 s}f(z)
-
{2 \pi \alpha_s F\over s}{s^2\over M_S^4} g(z)
+
{8\pi F^2 \over s} {s^4\over M_S^8}  h(z)
\right]
\label{difxs}
\end{eqnarray}

\noindent 
where
$z=p_1\cdot(p_4-p_3)/p_1\cdot p_2$ 
is the center of mass scattering angle
and $s=(p_1+p_2)^2$. In the
limit where the mass of the quarks is neglected, the formulas for $f(z)$,
$g(z)$ and $h(z)$ and $k_s$ are given in Table~1 
where the SM part agrees with the 
calculations given for example in~\cite{collider}.
Note that in cases
where there are two identical particles in the final state, a factor of
$1/2$ is included in $k_s$ so in all cases phase space should be
integrated over the range $-1\leq z \leq +1$.

\begin{table}
\begin{center}
{\large\bf Table 1}
\end{center}
\begin{tabular}{|c|c|c|}
\hline
Process & $k_s$ & f(z)  \\
\hline
\hline
$q \bar q \to q^\prime \bar q^\prime$
&
$1/36$
&
$8(1+z^2)$
\\
\hline
$q q^\prime \to q q^\prime$;
$q \bar q^\prime \to q \bar q^\prime$
&
$1/36$
&
$16{5+2z+z^2\over (1-z)^2}$
\\
\hline
$qq\to qq$
&
$1/72$
&
${32\over 3}{(z^2+11)(3z^2+1)\over (1-z^2)^2}$
\\
\hline
$q\bar q\to q\bar q$
&
$1/36$
&
${8\over 3}{(7-4z+z^2)(5+4z+3z^2)\over (1-z)^2}$
\\
\hline
$gg\to q\bar q$
&
$1/256$
&
${16\over 3}{ (9z^2+7)(1+z^2)\over 1-z^2 }$
\\

\hline
$gq\to gq$
&
$1/96$
&
${32\over 3}{(5+2z+z^2)(11+5z+2z^2)\over (1+z)(1-z)^2} $
\\
\hline
$gg\to gg$
&
$1/512$
&
$288{(3+z^2)^3\over (1-z^2)^2}$
\\
\hline
\multicolumn{3}{c}{ ~ }\\
\hline
Process &  g(z) & h(z) \\
\hline
\hline
$q \bar q \to q^\prime \bar q^\prime$
&
0
&
${9\over 256}(1-3z^2+4z^4)$
\\
\hline
$q q^\prime \to q q^\prime$;
$q \bar q^\prime \to q \bar q^\prime$
&
0
&
${9\over 2048}(149+232z+114z^2+16z^3+z^4)$
\\
\hline
$qq\to qq$
&
$-4{ 5-3z^2 \over  1-z^2 }$
&
${3\over 1024}(547+306z^2+3z^4)$
\\
\hline
$q\bar q\to q\bar q$
&
$-{1\over 4}{(11-14z-z^2)(1+z)^2\over  1-z }$
&
${3\over 2048}(443+692z+354z^2+116z^3+107z^4)$
\\
\hline
$gg\to q\bar q$
&
$-4(1+z^2)$
&
${3\over 8}(1-z^4)$
\\
\hline
$gq\to gq$
&
$2(5+2z+z^2)$
&
${3\over 8}(1+z)(5+2z+z^2)$
\\
\hline
$gg\to gg$
&
$120(3+z^2)$
&
${9\over 4}(3+z^2)^2$
\\ 
\hline
\end{tabular}
\caption{
In 
this table, we give the value of $k_s$ and the functions 
$f(z)$, $g(z)$ and $h(z)$ which define the differential cross section in 
eq.~(\ref{difxs}) for each of the $2\to 2$ processes relevant to dijet 
production in hadron collisions. 
The variable $z$ is the scattering angle in the center of mass frame 
given by $(t-u)/s$ and in
all cases the total cross section is given by integrating $z$ over the
range $-1\leq z\leq +1$.
}
\end{table}

The total differential two jet cross sections is shown in Fig.~1 at the
LHC $pp$ collider with $\sqrt{s_0}=14~TeV$, for $M_S=2$, $4$, $6$ and the
SM alone and at the Tevatron $p\bar p$ collider with $\sqrt{s_0}=2~TeV$,
for $M_S=0.75$, $1.5$ and the SM alone. Here, $s_0$ is the square of the
center of mass energy of the hadronic collision and $\tau=s/s_0$. In all
cases we have imposed the cut $|z|<0.5$ which tends to favor the graviton
scattering processes. The fraction of this differential cross section due
to various partonic subprocess for the LHC with $M_S=2~TeV$ and at the
Tevatron with $M_S=0.5~TeV$ is shown in Fig.~2(a) and 2(b). Of course, the
extrapolation of these curves beyond $M_S$ is not valid since at that
point new physical processes, such as the brane recoil effects
in~\cite{bando}, will enter and suppress the effect.  In Fig.~1 this point
is indicated by the black circles and so the portion of the curve to the
right of the circles may depend on the cutoff mechanism. In these results we
have used the CTEQ4M structure functions, set \#1~\cite{cteq}.

In the case of the LHC, one can see that the dominant contributions are
from $gg\to gg$ and $qg\to qg$ for $\tau<0.1$, which results from the
dominance of gluons for lower $\tau$. At $\tau>0.1$, $qq\to qq$ becomes
important due to the hard component of the structure functions of the
constituent quarks.  At the Tevatron, $gg\to gg$ and $qg\to qg$ are
dominant at low $\tau$ while here $q\bar q\to q\bar q$ will be dominant at
larger $\tau$.

In order to get an idea of what the reach of these signals are, we
consider imposing cuts of the form $\tau>\tau_0$ since, clearly, the SM
backgrounds are more important at lower $\tau$. In Fig.~3 we show the
maximum value of $M_S$ for which the difference between the Standard Model
and the Standard Model with gravitation has a $3\sigma$ significance both
at the LHC and the Tevatron.  In this graph we have taken an integrated
luminosity for the LHC of $30~fb^{-1}$ and for the Tevatron of
$2~fb^{-1}$. From this graph it is apparent that, for an optimal $\tau_0$
cut of $\sim 0.2$, the reach according to this criterion is $\sim 10~TeV$,
while, for the Tevatron at $\tau_0\sim 0.2$, the reach is $\sim 1.5~TeV$. 
A study~\cite{tev2jet} of existing $CDF$ and $D0$ two jet data gives a
bound of $M_S<1.2~TeV$.

A related process which has been previously considered~\cite{drellyan} is
the Drell-Yan process which at hadron colliders, $pp$ or $p\bar p\to
e^+e^-+X$. In that case, at a $\sqrt{s_0}=14~TeV$ LHC, with integrated
luminosity of $30~fb^{-1}$, one obtains a reach of about $5.6~TeV$, while
at the Tevatron with $\sqrt{s_0}=2~TeV$, given an integrated luminosity of
$2~fb^{-1}$ one obtains a reach of $1.3~TeV$.

It is interesting to compare these two jet results to those which may be
obtained at the NLC by studying $2\to 2$ scattering processes.  Many such
processes have been considered in the literature~\cite{drellyan,
nlc_tests, nlc_moler}.  In particular it was pointed out
in~\cite{nlc_moler} that the $e^-e^-\to e^-e^-$ mode does somewhat better
than the $e^+e^-$ modes at the same luminosity.  For the sake of
definiteness, let us consider the reach of a $e^+ e^-$ or $e^-e^-$
collider with $\sqrt{s}=1~TeV$ and integrated luminosity of $100~fb^{-1}$,
where we impose a cut on the two final state particles of $|z|<0.5$. In
this case we find that the reach in $M_S$ is $7~TeV$ for $e^+e^-\to
\mu^+\mu^-$, $4~TeV$ for $e^+e^-\to 2~jets$, $5.5~TeV$ for $e^+e^-\to
\gamma\gamma$, $8.5~TeV$ for $e^+e^-\to e^+e^-$ and $9.2~TeV$ for
$e^-e^-\to e^-e^-$.

Another proposed mode of operation of an NLC is to convert it into a gamma
gamma collider by scattering optical frequency laser beams off of the
electron beams~\cite{ginzberg}.  This allows, for instance the study of
$\gamma\gamma\to \gamma\gamma$, where there is no tree level SM
background. The leading SM contributions is given by the box diagram
derived in~\cite{boxgraph}. These processes were studied extensively
in~\cite{hooman, cheung} where in~\cite{hooman} detailed consideration is
given to optimization of the cuts and polarization of the photons and the
electrons. A reach of $3.5~TeV$ is thus obtained for $n=6$ and likewise
$3.8~TeV$ for $n=4$ based on an NLC with electron-positron center of mass
energy $\sqrt{s_{ee}}= 1~TeV$.  Of course one may also consider a NLC
where only one of the electron beams is converted into a photon beam. At
such a collider, one may study $e^\pm\gamma\to e^\pm\gamma$. For this
process a reach of $M_S\sim 7.5~TeV$ is found~\cite{hooman2}, again for
$n=4$ based on an NLC with electron-positron center of mass energy
$\sqrt{s_{ee}}= 1~TeV$ and an integrated luminosity of $100~fb^{-1}$.

The case of two photons going to two jets, $\gamma\gamma\to q\bar q$ and
$\gamma\gamma\to gg$, has been considered in detail in
\cite{gammagammajj}. They find that in a $\gamma\gamma$ collider based on
a $500~GeV$ electron-positron machine, the sensitivity is $(3.2,2.8)~TeV$
for $n=(4,6)$ while the sensitivity is $(11.1,9.4)~TeV$ at a $2~TeV$
machine.

In conclusion then, two jet signals at the LHC can give a reach of about
$10~TeV$ for $M_S$ which is quite favorable to the reaches obtainable via
Drell-Yan ($5.8~TeV$) and monojet signals (i.e. $4.5$; $3.3~TeV$ for
$n=2$; $6$). An NLC collider running in $e^-e^-$ mode could achieve
comparable reaches i.e., $8.5~TeV$, however it is unclear if such a
collider would run extensively in this mode. In $e^+e^-$ mode, of the
processes considered, $e^+e^-$ gives the best reach of $6.8~TeV$.  Even
though there are large SM backgrounds to the dijet cross section at
hadronic colliders, the fact that graviton exchange dominantly contributes
only at the highest values of $\tau$ makes this signal viable.

\medskip
\flushleft{\bf Acknowledgments}
This research was supported in part by US DOE Contract Nos.  DE-FG02-94ER40817
(ISU) and DE-AC02-98CH10886 (BNL)

\newpage

\newpage

\begin{center}
{\large\bf Figure Captions}
\end{center}

{\bf Figure 1} The total differential cross sections $d\sigma/d\tau$ are
shown as a function of $\tau$ for $n=4$ where the acceptance cut $|z|\leq
0.5$ has been imposed for various values of $M_S$. Solid lines represent
the contribution at the LHC 
($\sqrt{s_0}=14~TeV$)
if $M_S=2~TeV$ (upper solid line), $4~TeV$,
$6~TeV$ and the Standard Model alone (lower solid line). The dashed lines
represent the contributions at the Tevatron 
($\sqrt{s_0}=2~TeV$)
if $M_S=0.75~TeV$ (upper
dashed line), $1.5~TeV$ and the Standard Model alone (lower dashed line).
The circles indicate where $M_S^2=\tau s_0$.

\medskip

{\bf Figure 2}
$(d\sigma_i/d\tau)/(d\sigma/d\tau)$ 
as a function of $\tau$
for each partonic mode
with $n=4$ is
shown; in Fig.~2(a)
the LHC is considered with $pp$ collisions at $\sqrt{s_0}=14~TeV$ taking
$M_S=2~TeV$ while in Fig.~2(b)  the results for the Tevatron is
considered with $p\bar p$ collisions at $\sqrt{s_0}=2~TeV$ taking
$M_S=0.5~TeV$. In both cases a cut of $z<0.5$ is imposed. The
subprocesses are 
$q\bar q \to q^\prime \bar q^\prime$ (thin dashed line); 
$q q^\prime \to q q^\prime$ (thin dotted line);  
$q \bar q^\prime \to q\bar q^\prime$ (thick dot dash line);  
$q q \to q q$ (thin dot dash line); 
$q \bar q \to q \bar q$ (thick dotted line);  
$gg \to q \bar q$ (thick long dashed line);  
$ q \bar q \to gg $ (thin solid line);  
$ qg \to qg +\bar q g\to \bar q g$ (thick dashed line);  
$ g g \to gg $ (thick solid line).

\medskip

{\bf Figure 3} The reach of the Tevatron and LHC in the case of $n=4$ as a
function of a lower cut in $\tau$ based on the total cross section as in
Fig.~1. In both cases a criterion of 3 sigma was used. In the LHC case an
integrated luminosity of $30~fb^{-1}$ was assumed while in the case of the
Tevatron an integrated luminosity of $2~fb^{-1}$ was assumed.

\newpage

\begin{center}
{\large\bf Figure 1}
\end{center}

\begin{figure}[htb]
\begin{center}
\epsfig{file=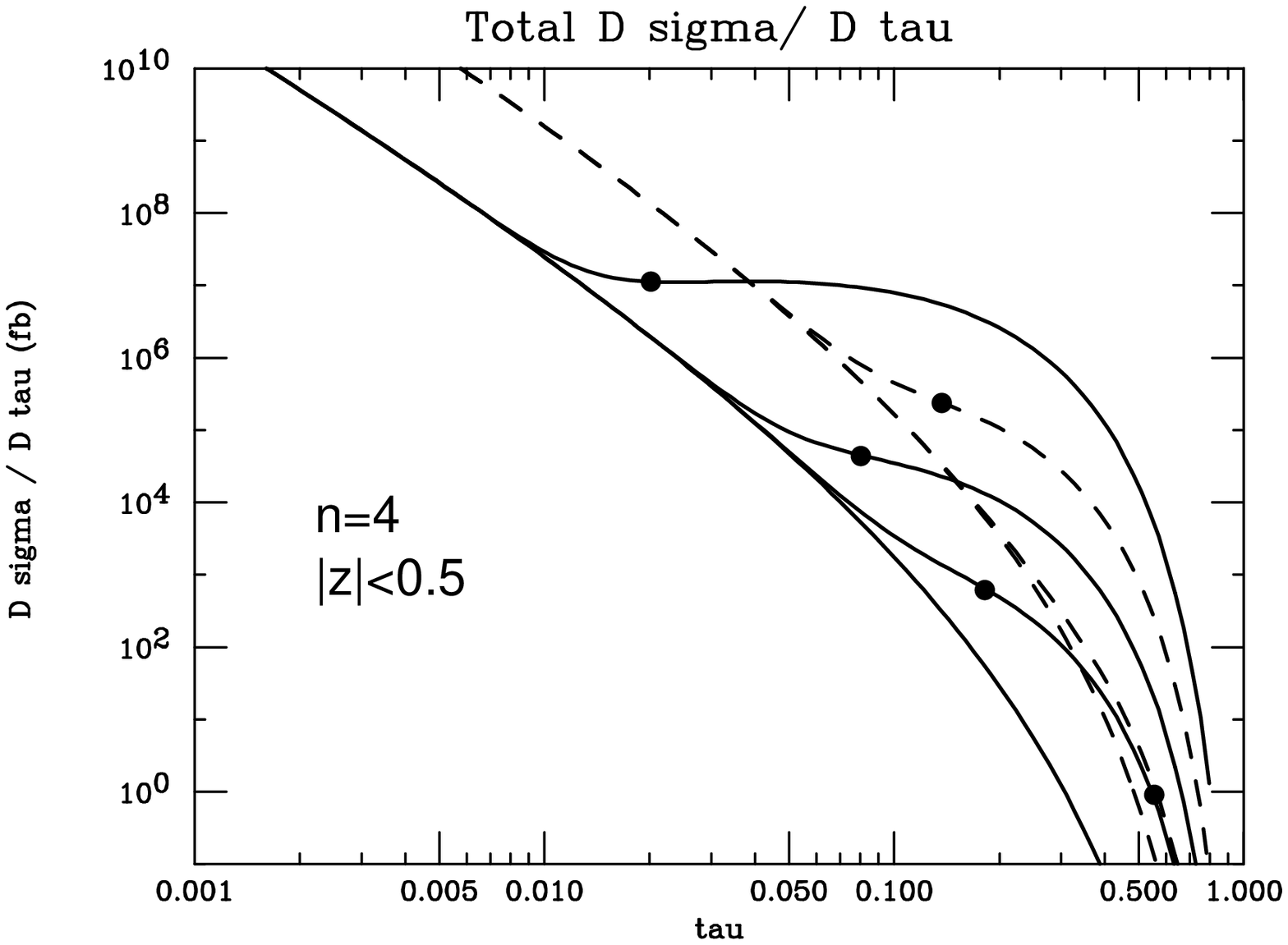,width=5 in}
\end{center}
\end{figure}

\newpage

\begin{center}
{\large\bf Figure 2(a)}
\end{center}

\begin{figure}[htb]
\begin{center}
\epsfig{file=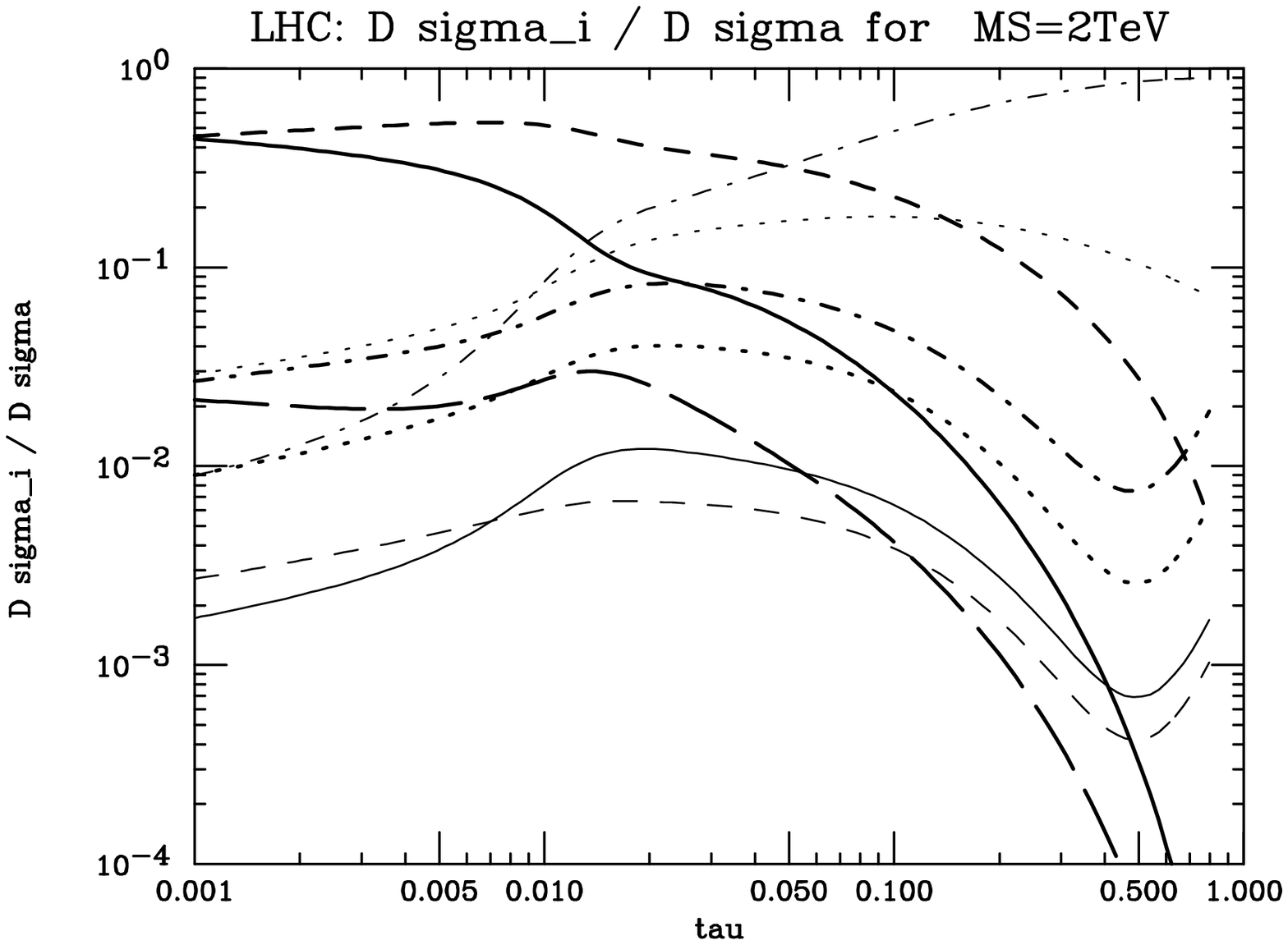,width=5 in}
\end{center}
\end{figure}

\newpage

\begin{center}
{\large\bf Figure 2(b)}
\end{center}

\begin{figure}[htb]
\begin{center}
\epsfig{file=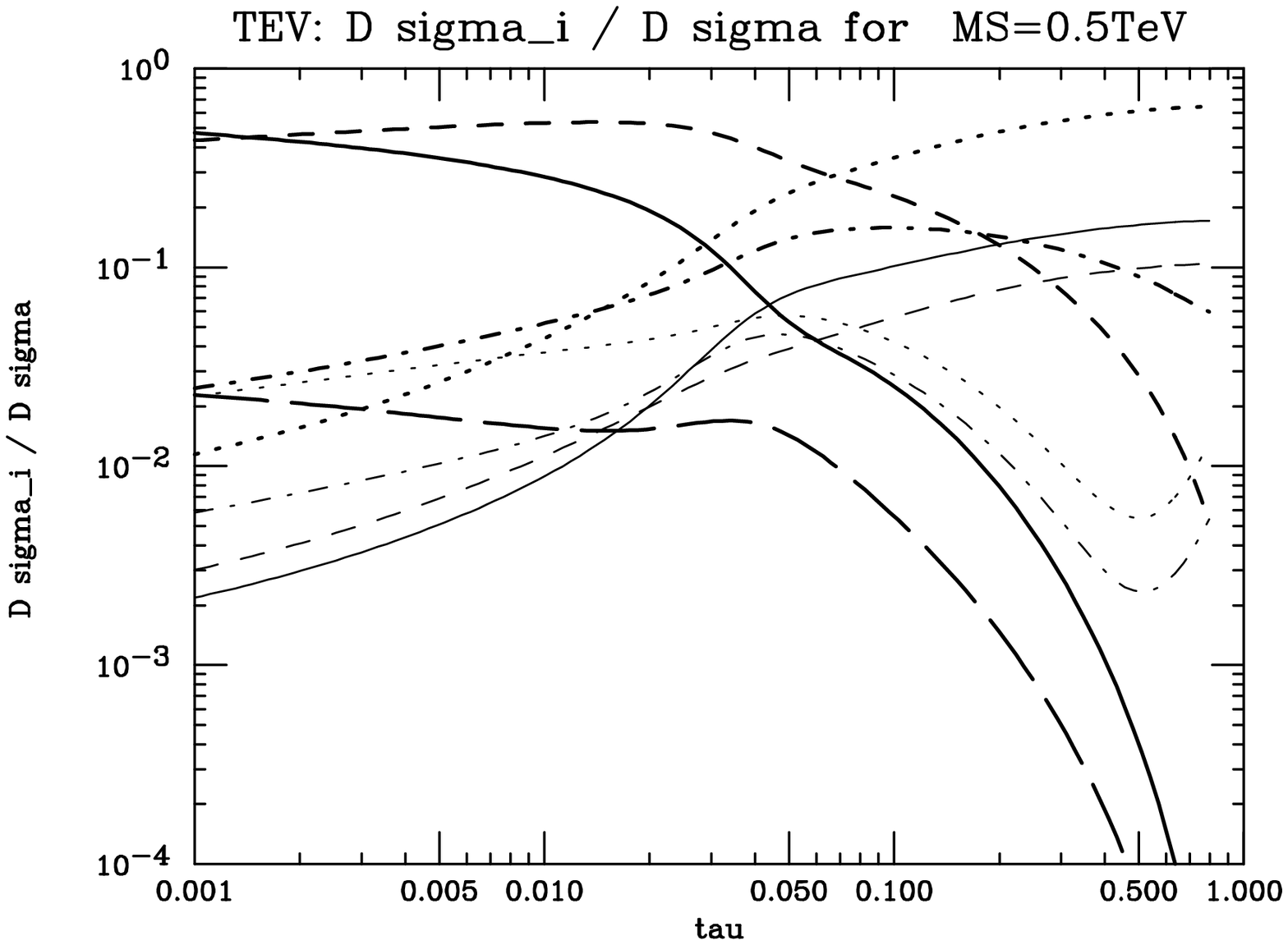,width=5 in}
\end{center}
\end{figure}

\newpage

\begin{center}
{\large\bf Figure 3}
\end{center}

\begin{figure}[htb]
\begin{center}
\epsfig{file=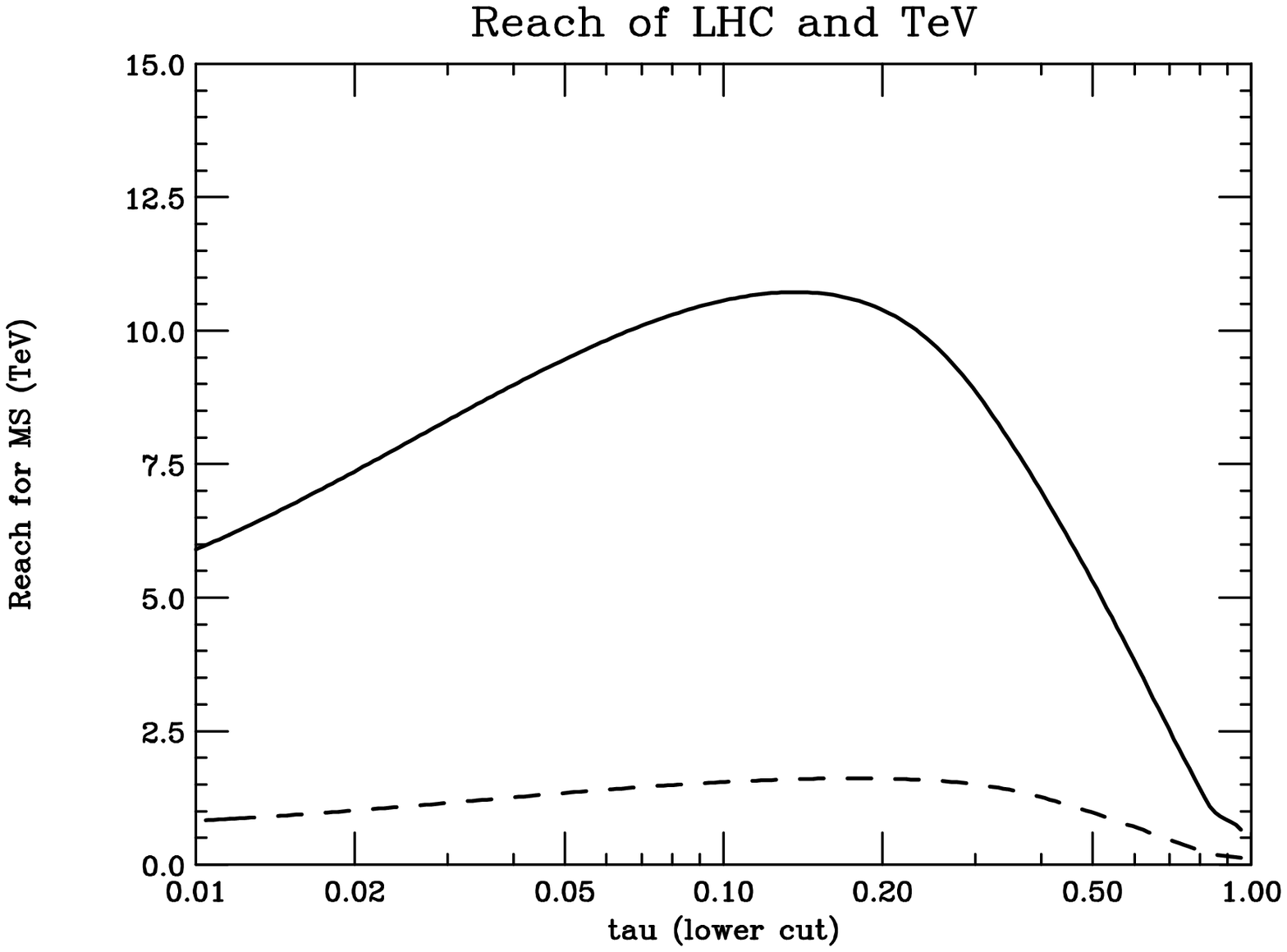,width=5 in}
\end{center}
\end{figure}

\end{document}